\documentclass[aps,prl,10pt,superscriptaddress,twocolumn,longbibliography]{revtex4-2}

\usepackage{graphicx}
\usepackage{xcolor}
\usepackage{hyperref}
\usepackage{multirow}

\hypersetup{colorlinks=true, linkcolor=blue, urlcolor=blue, citecolor=blue}

\begin{document}

\title{Deep Learning Based Recalibration of SDSS and DESI BAO Alleviates Hubble and Clustering Tensions}

\author{Rahul Shah}
\email{rahul.shah.13.97@gmail.com (corresponding author)}
\thanks{\\These authors contributed equally to this work}
\affiliation{Physics and Applied Mathematics Unit, Indian Statistical Institute,\\203, B.T. Road, Kolkata 700 108, West Bengal, India}

\author{Purba Mukherjee}
\email{purba16@gmail.com}
\thanks{\\These authors contributed equally to this work}
\affiliation{Centre for Theoretical Physics, Jamia Millia Islamia,\\ New Delhi 110025, Delhi, India}

\author{Soumadeep Saha}
\email{soumadeep.saha97@gmail.com}
\affiliation{Computer Vision and Pattern Recognition Unit, Indian Statistical Institute,\\203, B.T. Road, Kolkata 700 108, India}

\author{Utpal Garain}
\email{utpal@isical.ac.in}
\affiliation{Computer Vision and Pattern Recognition Unit, Indian Statistical Institute,\\203, B.T. Road, Kolkata 700 108, India}

\author{Supratik Pal}
\email{supratik@isical.ac.in}
\affiliation{Physics and Applied Mathematics Unit, Indian Statistical Institute,\\203, B.T. Road, Kolkata 700 108, West Bengal, India}

\begin{abstract}
Conventional calibration of Baryon Acoustic Oscillations (BAO) data relies on estimation of the sound horizon at drag epoch $r_d$ from early universe observations by assuming a cosmological model. We present a recalibration of two independent BAO datasets, SDSS and DESI, by employing deep learning techniques for model-independent estimation of $r_d$, and explore the impacts on $\Lambda$CDM cosmological parameters. Significant reductions in both Hubble ($H_0$) and clustering ($S_8$) tensions are observed for both the recalibrated datasets. Moderate shifts in some other parameters hint towards further exploration of such data-driven approaches.
\end{abstract}

\maketitle

\textit{Introduction.}--One of the major objectives of modern cosmology is to build up a synergy of multiple observational datasets to constrain cosmological parameters. Baryon Acoustic Oscillations (BAO), Cosmic Microwave Background (CMB), and Type-Ia Supernovae (SNIa) observations form the combined baseline dataset along with the local measurement of the Hubble constant (SH0ES) \cite{Riess:2021jrx}. However, the growing discordances in the values of key cosmological parameters, particularly the Hubble constant tension \cite{DiValentino:2021izs} (a persistent discrepancy in $H_0$, a measure of the current expansion rate of the universe) and the clustering tension \cite{DiValentino:2020vvd} (involving $S_8$, that is related to the amplitude of matter fluctuations at $8h^{-1}$ Mpc scale $\sigma_{8,0}$, a measure of clumpiness of the universe) have sparked many debates. Potential reasons for these include \cite{Abdalla:2022yfr}: (1) systematic errors in one or more experiments, (2) physics beyond the baseline $\Lambda$ Cold Dark Matter ($\Lambda$CDM) model, and (3) correlated biases and miscalibration between datasets. While a gross systematic error is very unlikely to happen for any of the present missions, and extensive work on beyond-$\Lambda$CDM models has been in progress for the last few years with nominal to moderate success \cite{DiValentino:2021izs}, not much investigation has been made or promised on the third possibility till date.

A prominent example of such bias arises in the BAO measurements, which encode the primordial sound horizon at the baryon drag epoch $r_d$ within the large-scale structure. They act as a \textit{standard ruler}, offering critical insights into the background dynamics and evolution of perturbations \cite{Bassett:2009mm}. However, their cosmological utility depends on an accurate calibration of $r_d$, which is traditionally derived from early-universe (CMB-based) observations under the assumption of the $\Lambda$CDM model \cite{Eisenstein:1997ik}, defined as,
\begin{equation}
    r_d = \int_{z_{\text{drag}}}^\infty \frac{c_s(z)}{H(z)} \, {\rm d}z \, ,
\end{equation} where $z_{\text{drag}}$ corresponds to the epoch when photon drag on baryons becomes negligible, $c_s(z)$ is the sound speed in the photon-baryon fluid and $H(z)$ is the Hubble parameter for a corresponding cosmological model. 

This widely-adopted \textit{standard calibration} hence raises an important question: does the calibration of BAO data using CMB-derived $r_d$ introduce bias into the dataset and hence potentially obscure their true cosmological implications? In order to investigate this from a fresh perspective, we revisit the calibration of BAO datasets, SDSS-IV \cite{eBOSS:2020yzd} and DESI DR1 \cite{DESI:2024mwx}, using data-driven methods. We employ a machine learning approach, by utilizing the deep learning suite \texttt{LADDER} \cite{Shah:2024slr}, to reconstruct the Hubble diagram from Pantheon SNIa \cite{Pan-STARRS1:2017jku} data, and interpolate it at BAO redshifts in order to infer $r_d$. Our analysis only assumes an astrophysical prior on the absolute magnitude $M_B$ of SNIa, thus avoiding reliance on the underlying cosmological model. 

\begin{figure}[!ht]
    \centering
    \includegraphics[width=0.48\textwidth]{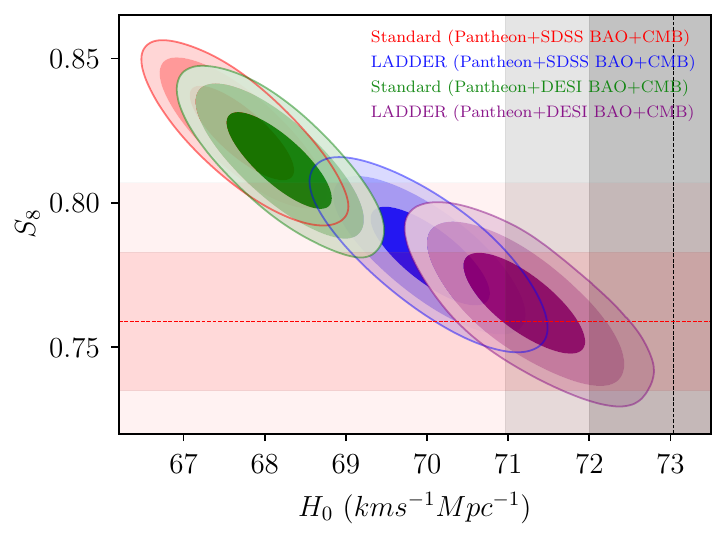}
    \caption{Comparison of $H_0$-$S_8$ constraints for different combinations of calibration and datasets. The gray and red shaded regions correspond to the late-time measurements of $H_0=73.04\pm1.04 \text{ km s}^{-1}\text{Mpc}^{-1}$ \cite{Riess:2021jrx} and $S_8 = 0.759\pm0.024$ \cite{DES:2021bvc}.} 
    \label{fig:H0S8}
\end{figure}

This \textit{letter} further explores the implications of this alternative calibration of BAO on cosmological constraints. We find that the two most crucial tensions, \textit{i.e.}, those in $H_0$ and $S_8$, are alleviated to a considerable extent, as can be seen in Fig. \ref{fig:H0S8}. This figure alone serves as our major result. This, along with slight to moderate shifts in some other parameters, might warrant a revisiting of the CMB dataset to ensure its role as an unbiased calibration tool and might call for further data-driven approaches with potentially interesting outcomes. Moreover, both SDSS and DESI BAO show similar trends, with DESI performing better with improved precision, thereby validating the robustness of our analysis.

\squeezetable
\begin{table}[!ht]
    \begin{ruledtabular}
        \resizebox{0.5\textwidth}{!}{\renewcommand{\arraystretch}{2.0} \setlength{\tabcolsep}{3 pt}
            \begin{tabular}{c c c}
            \textbf{Parameters} & \textbf{SDSS} & \textbf{DESI} \\
            \colrule
            {\boldmath$r_d$} & $137.651^{+2.157}_{-2.153}$ & $137.586^{+2.154}_{-2.151}$ \\
            \end{tabular}
        }
    \end{ruledtabular}
    \caption{Values of $r_d$ (in Mpc) from \texttt{LADDER} calibration of BAO data.}
    \label{tab:mbrd}
\end{table}

\squeezetable
\begin{table*}[!ht]
    \begin{ruledtabular}
    \resizebox{1.0\textwidth}{!}{\renewcommand{\arraystretch}{1.25} \setlength{\tabcolsep}{5 pt}
        \begin{tabular}{c c c c c}
        \multirow{2}{*}{\textbf{Parameters}} & \multicolumn{2}{c}{\textbf{Standard calibration}} & \multicolumn{2}{c}{\textbf{\texttt{LADDER} calibration}} \\
        \cline{2-3}
        \cline{4-5}
        & \textbf{SDSS} & \textbf{DESI} & \textbf{SDSS} & \textbf{DESI}\\ 
        \hline
        {\boldmath$100{\Omega_b}{h^2}$} & $2.244\pm0.013$ & $2.251\pm0.014$ & $2.296\pm0.014$ & $2.330\pm0.014$\\
        {\boldmath${\Omega_c}{h^2}$} & $0.11928\pm0.00088$ & $0.11828\pm0.00088$ & $0.11480\pm0.00099$ & $0.11292\pm0.00099$\\
        {\boldmath$100{\theta_s}$} & $1.04193\pm0.00028$ & $1.04205\pm0.00029$ & $1.04258\pm0.00028$ & $1.04299\pm0.00028$\\
        {\boldmath${\ln{\left({10^{10}A_s}\right)}}$} & $3.047\pm0.014$ & $3.051\pm0.015$ & $3.071\pm0.018$ & $3.089\pm0.019$\\
        {\boldmath$n_s$} & $0.9673\pm0.0037$ & $0.9698\pm0.0036$ & $0.9793\pm0.0039$ & $0.9852\pm0.0040$\\
        {\boldmath${\tau}$} & $0.0560\pm0.0072$ & $0.0591\pm0.0075$ & $0.0718\pm0.0092$ & $0.0816\pm0.0096$\\
        \\
        {\boldmath$M_B$} & $-19.417\pm0.011$ & $-19.405\pm0.011$ & $-19.356\pm0.012$ & $-19.326\pm0.012$\\
        {\boldmath$H_0$} & $67.72\pm0.40$ & $68.18\pm0.40$ & $70.04\pm0.46$ & $71.20\pm0.47$\\
        {\boldmath$\Omega_m$} & $0.3105\pm0.0053$ & $0.3043\pm0.0052$ & $0.2822\pm0.0055$ & $0.2700\pm0.0053$\\
        {\boldmath$\sigma_{8,0}$} & $0.8101\pm0.0060$ & $0.8089\pm0.0063$ & $0.8060\pm0.0069$ & $0.8066\pm0.0074$\\
        {\boldmath$S_8$} & $0.824\pm0.010$ & $0.815\pm0.010$ & $0.782\pm0.011$ & $0.765\pm0.011$\\
        \end{tabular}
    }
    \end{ruledtabular}
\caption{Comparison between constraints, employing standard CMB-based calibration \textit{vs} \texttt{LADDER} calibration of BAO, in combination with Planck+Pantheon. $H_0$ is in units of $\text{km s}^{-1}\text{Mpc}^{-1}$.}
\label{tab:bao_results}
\end{table*}

\textit{Methodology.}--Our analysis consists of three steps: (1) model-independent recalibration of BAO, (2) constraining the $\Lambda$CDM model with this new BAO dataset in combination with CMB and SNIa, and (3) comparing the constraints between the standard \textit{vs} \texttt{LADDER} calibrations. We begin by employing the publicly available \texttt{LADDER} deep learning suite \cite{laddercode}, and follow the methodology described in Section 4.2 of \cite{Shah:2024slr}. This involves training a neural network on the apparent magnitude data from the Pantheon SNIa compilation, making predictions at BAO redshifts and hence inferring the value of $r_d$ in a cosmological model-independent manner, by employing Markov chain Monte Carlo (MCMC) methods (using \texttt{emcee} \cite{Foreman-Mackey:2012any}). The only assumption made is an astrophysical prior on the absolute magnitude of SNIa in the B-band ($M_B=-19.2478\pm0.0294$ \cite{Pan-STARRS1:2017jku}), which is necessary in this framework, since without it $r_d$ remains unconstrained. The inferred $r_d$ (Table \ref{tab:mbrd}) is then used to recalibrate two independent BAO compilations from SDSS-IV and DESI DR1 surveys, by collecting them from Tables III \& I of \cite{eBOSS:2020yzd} and \cite{DESI:2024mwx} respectively. 

\begin{figure*}[!ht]
    \centering
    \includegraphics[width=\textwidth]{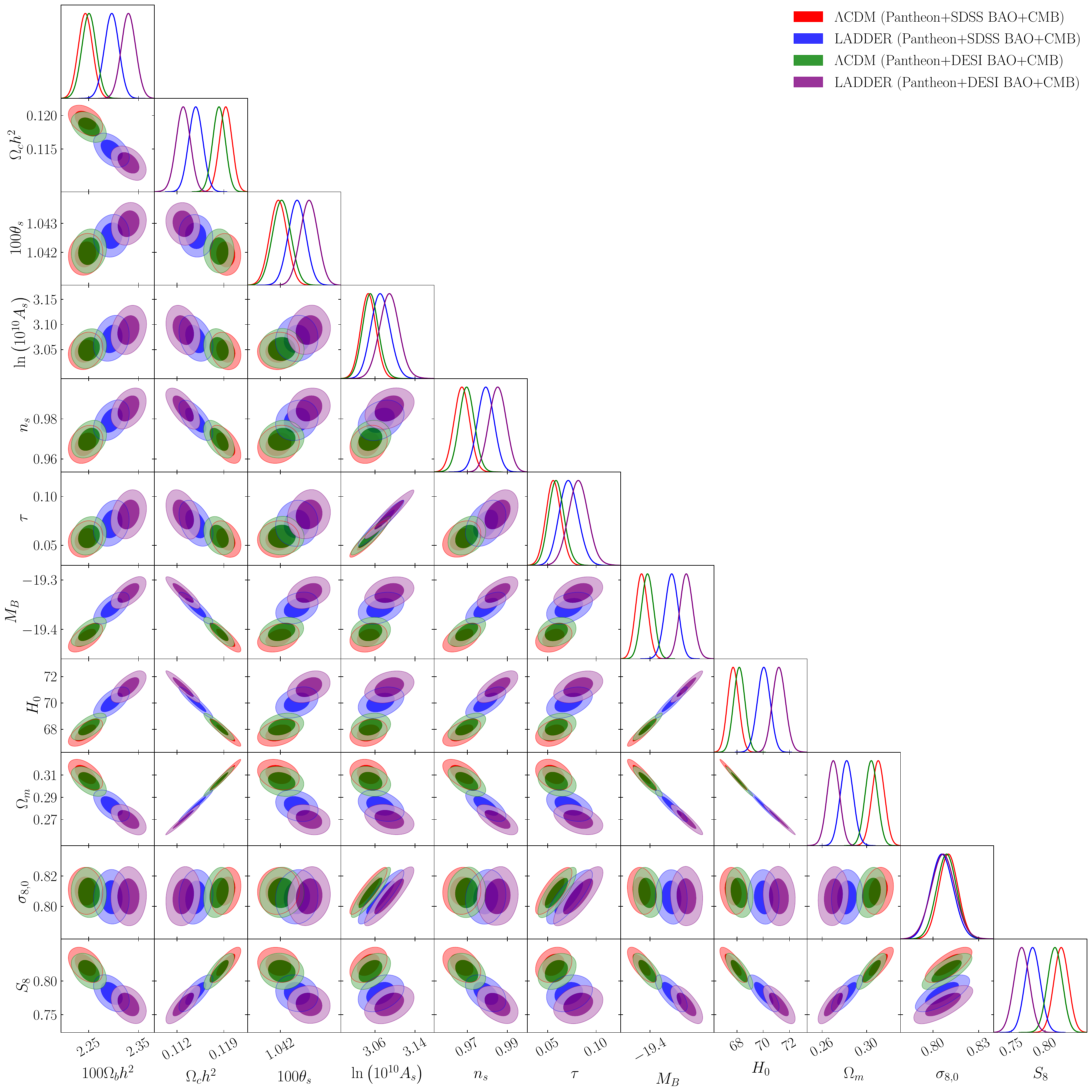}
    \caption{Comparison of constraints for different combinations of calibration and datasets. $H_0$ is in units of $\text{km s}^{-1} \text{ Mpc}^{-1}$.}
    \label{fig:triangle}
\end{figure*}

Next, we constrain the $\Lambda$CDM model parameters with a combination of Planck 2018 TTTEEE + low l + low E + lensing \cite{Planck:2019nip,Planck:2018vyg,Planck:2018lbu}, BAO and Pantheon \cite{Pan-STARRS1:2017jku} (using \texttt{CLASS} \cite{Lesgourgues:2011re,Blas:2011rf} and \texttt{MontePython} \cite{Brinckmann:2018cvx,Audren:2012wb}). To draw direct comparisons, we obtain constraints for both the standard and \texttt{LADDER} calibrated BAO datasets. For this purpose, we modify the default \texttt{MontePython} likelihood files for the standard-calibrated BAO to impose the \texttt{LADDER}-calibrated $r_d$ on the BAO datasets. The comparative constraints thus obtained are presented in Table \ref{tab:bao_results} and Figs. \ref{fig:H0S8} \& \ref{fig:triangle} (generated using \texttt{GetDist} \cite{Lewis:2019xzd}).

\textit{Results \& Discussion.}--As already mentioned, in the standard method, $r_d$ is derived from early-time physics through CMB and extrapolated to BAO redshifts using the vanilla $\Lambda$CDM model, thereby imposing potential biases from both the CMB and the $\Lambda$CDM model on BAO. In contrast, our \texttt{LADDER} approach builds a distance ladder from low- to high-redshift SNIa observations and uses deep learning to interpolate between the BAO redshifts. This method thus helps overcome both the model- and CMB-bias on BAO datasets, which in turn enables cosmological parameter inference from BAO observations independent of early-universe assumptions.

Under the $\Lambda$CDM assumption, the CMB yields $r_d=147.09\pm0.26~\mathrm{Mpc}$ \cite{Planck:2018vyg}. In contrast, our model-independent determination of $r_d$ from the \texttt{LADDER} framework is notably lower (Table \ref{tab:mbrd}), which reflects its dependence on the astrophysical prior for $M_B$. Since BAO effectively constrains the product $H_0r_d$, the impact of a lower value for $r_d$, as obtained via this recalibration, will be revealed shortly.

There is statistically no significant difference between constraints corresponding to the two BAO datasets (SDSS and DESI) when calibrated with the conventional method. However, our recalibration of $r_d$ fundamentally impacts the constraints on a few cosmological parameters, by allowing moderate to significant shifts in their estimates compared to the standard CMB-based calibration. As shown in Fig. \ref{fig:triangle}, these shifts appear for both the BAO datasets, albeit with more pronounced effects for DESI. Moreover, such identical trends across the two independent BAO datasets indicate the robustness of the \texttt{LADDER} calibration method against BAO systematics. Key shifts thus induced include:

(1) $M_B$ is kept as a free parameter in the model constraining stage. It attains a value partway between the injected astrophysical prior at the \texttt{LADDER} calibration stage and that obtained from standard calibration. \\

(2) An increase in $H_0$ from the \texttt{LADDER}-calibrated $r_d$ is driven by the larger angular size of the sound horizon at the last scattering surface $\theta_s$. This is reflected as a reduced angular diameter distance $D_A$ to the last scattering surface, requiring a faster expansion rate to match the observed angular scale. This brings $H_0$ closer to local measurements, such as the SH0ES value \cite{Riess:2021jrx}, thereby helping to alleviate the $H_0$ tension.

(3) There is a reduction in the present-day matter density $\Omega_m$ as $H_0$ increases, which reflects the geometric degeneracy between these parameters. In cosmological models, an increase in $H_0$ can be offset by a decrease in $\Omega_m$, as both parameters influence $D_A$. The change in the expansion rate is balanced with the reduced $\Omega_m$, thereby maintaining consistency with the observed data.

(4) $\sigma_{8,0}$ remains unaffected because it depends on late-time structure growth which is governed by $\Omega_m$ and $H_0$. While recalibration alters the latter two parameters, affecting the early universe's expansion, it does not significantly change the growth of structure at later times.

(5) Even though $\sigma_{8,0}$ remains more or less at the baseline value, a reduction in the matter density $\Omega_{m}$ shifts $S_8$ towards values consistent with estimates from weak lensing and galaxy cluster surveys that probe structure growth at later epochs, thereby tending to relax the clustering tension (see, \textit{e.g.}, the references in \cite{Shah:2024rme}). 

(6) Notably, the combined analysis of the recalibrated BAO data with CMB and SNIa impacts the estimation of some of the parameters, for which CMB values were assumed to be sacrosanct, from mild to moderate extents. There is an increase in the baryon density $\Omega_bh^2$ with a corresponding decrease in the inferred cold dark matter density $\Omega_ch^2$. Further, slight shifts in the primordial parameters $A_s$ and $n_s$ seem to suggest slightly more initial density fluctuations and a flatter power spectrum than Planck estimates. Meanwhile, the increase in the reionization optical depth $\tau$ points towards a slightly earlier and/or more extended reionization epoch, which may alter the interpretation of CMB data and its interaction with the ionized gas \cite{Weiland:2018kon,Paoletti:2020ndu}. While we do not claim that these differences are alarming, this might call for cross-checks with other CMB missions with potentially different pivot scales, as mentioned later on in this \textit{letter}.

It is worth noting that \texttt{LADDER} calibration induces a significant \textit{in-plane} shift in the $H_0$-$S_8$ parameter space (Fig. \ref{fig:H0S8}). This considerably alleviates both tensions while preserving their mutual correlation. In our opinion, this is one of the most crucial results of the present analysis, which can help overcome the existing notion that $H_0$ and $S_8$ tensions may not be simultaneously resolved using current datasets \cite{Bhattacharyya:2018fwb, DiValentino:2020kha}. Our analysis suggests that even though the statement above is partially true, it is based upon the standard CMB-based BAO calibration technique. Thus, the \texttt{LADDER}-based model-independent recalibration of BAO helps overcome this limitation in a novel way. The observed \textit{in-plane} shift has been reported in earlier works which compared distance ladder and inverse distance ladder methods and investigated how calibrating cosmological parameters with late- \textit{vs} early-time anchors can systematically influence parameter spaces \cite{eBOSS:2020yzd,Perivolaropoulos:2024yxv}. Our analysis thus implies that resolving the tensions between $H_0$ and $S_8$ may be better achieved by recalibrating parameters rather than modifying the underlying cosmological model, thus preserving the baseline $\Lambda$CDM framework.

While both $H_0$ and $S_8$ shift in the desired directions, it is important to be alert towards mild to moderate shifts in other parameters. A slight preference for a higher value of $\tau$ is consistent with findings from other CMB analyses \cite{Weiland:2018kon,Giare:2023ejv} and slight shifts in $A_s$ and $n_s$ could be intriguing for the following reason. Constraints on these parameters, whether obtained solely from CMB or from a combination of CMB and CMB-calibrated BAO, appear consistent \cite{Planck:2018vyg}. This consistency, however, is expected as BAO data is naturally aligned to CMB as a result of conventional calibration. Rather, it raises the question of whether such an alignment could introduce an implicit bias in the BAO dataset. We hence propose that one should employ data-driven recalibration techniques, whenever possible, to arrive at model-agnostic constraints and help isolate potential systematics in individual datasets, or those arising from potentially biased calibration. 

These results would strongly advocate for a non-CMB-based, model-independent calibration of BAO as a robust approach to constraining cosmological parameters. However, some concerns surface when examining the integrity of resulting constraints from the combination of recalibrated BAO, CMB, and SNIa datasets. This is echoed by the tendency of CMB data to drag the value of $M_B$ \cite{Camarena:2021jlr} away from the injected astrophysical prior in the \texttt{LADDER} pipeline, complicating the interpretation of joint analyses. Should there be any hidden systematics in the CMB data, our recalibration ensures that these issues do not propagate to the BAO dataset. Furthermore, since BAO and Pantheon SNIa data are measured at the same redshift range, both datasets are subject to similar environmental conditions. As a result, any systematics affecting one dataset are likely to impact both in a comparable manner, reducing the potential for discrepancies between them. However, the shifts observed in certain parameters indicate that it may be worthwhile to revisit the CMB data and reassess its role in the calibration of BAO.

\textit{Conclusions \& Outlook.}--In this \textit{letter} our primary goal is to highlight the inherent bias introduced by the standard BAO calibration method from CMB data, assuming the $\Lambda$CDM model, and hence, the requirement of model-independent recalibration as a suitable alternative. We substitute the conventional, CMB-derived model-dependent $r_d$ measurement with a model-independent estimate, which only relies on an astrophysical prior for $M_B$. Through this, we effectively disentangle the BAO datasets from their dependence on CMB-based calibration. Moreover, the use of deep learning-based \texttt{LADDER} recalibration circumvents such biases by learning directly from the data. Our confidence in such a data-driven approach is bolstered by the \textit{in-plane} shift in the $H_0$-$S_8$ parameter space resulting in the simultaneous alleviation of tensions in $H_0$ and $S_8$ while remaining within the $\Lambda$CDM framework. 

The mild to moderate shifts in some other parameters should better be envisioned either as a limitation of the usefulness of CMB-data as a calibration tool for BAO or as the need for further model-independent recalibration techniques using novel deep learning algorithms. This calls for a thorough investigation in both these directions. The first approach can be taken up by cross-checking with current CMB missions other than Planck, with potentially different pivot scales, such as ACT \cite{ACT:2023kun} and SPT \cite{SPT:2004qip}, and upcoming missions such as CMB-S4 \cite{CMB-S4:2016ple}, LiteBIRD \cite{LiteBIRD:2020khw}, PICO \cite{NASAPICO:2019thw}, the Simons Observatory \cite{SimonsObservatory:2018koc}, \textit{etc}. For the second route, we acknowledge that \texttt{LADDER} is not necessarily the definitive tool for data-driven recalibration, but it does demonstrate the effectiveness of such methods. While \texttt{LADDER} shows promise as a starting point for future innovations in recalibration, it calls for direct extrapolation up to CMB redshifts for a unified understanding across all datasets. This \textit{letter} encourages further exploration into different machine learning techniques and data-driven approaches to improve upon cosmological inferences and refine conventional calibration practices.

Finally, we stress on the importance of ensuring the use of unbiased datasets when conducting cosmological analyses and constraining the benchmark $\Lambda$CDM, as well as other (beyond-$\Lambda$CDM) models. Maintaining independence among datasets is essential to reduce the effect of potentially uncorrected systematics and hence improve the robustness of the resulting conclusions. 


\textit{Data availability.}--The datasets used in this work are publicly available. The modified codes used for this study may be made available upon reasonable request.

\textit{Acknowledgments.}--RS acknowledges financial support from ISI Kolkata as a Senior Research Fellow. PM acknowledges financial support from the Anusandhan National Research Foundation (ANRF), Govt. of India under the National Post-Doctoral Fellowship (N-PDF File no. PDF/2023/001986). UG thanks the Indo-French Centre for the Promotion of Advanced Research (IFCPAR/CEFIPRA) for partial support through CSRP Project No. 6702-2. SP thanks the ANRF, Govt. of India for partial support through Project No. CRG/2023/003984. We acknowledge the use of the Pegasus cluster of the high performance computing (HPC) facility at IUCAA, Pune, India.

\bibliography{biblio}

\end{document}